# Statistical Performance Analysis of the MUSIC Algorithm in Angular Sectors

Antonios Bassias[1], Anthony Chronopoulos[2]

[1]Hellenic Telecommunications Organization (OTE)
99 Kifissias Ave., 151 24 Maroussi, Greece
E-mail: a.bassias@ote.gr
[2]Department of Computer Science, University of Texas, San Antonio
6900 North Loop 1604 West, San Antonio, Texas 78249-0667
E-mail: atc@cs.utsa.edu

Abstract This article deals with the problem of the statistical performance analysis of the MUSIC (Multiple Signal Classification) algorithm which is an eigen-decomposition based method for the estimation of the angles of arrival of signals received by an array of sensors. In past work the performance of the MUSIC algorithm was studied (via an asymptotic statistical analysis of the algorithm's null spectrum) for the case of two plane waves of equal power in noise. In this article, a new theoretical formula is derived for the signal to noise ratio resolution threshold of two uncorrelated, narrowband plane waves with equal powers in angular sectors received by an array of sensors. The accuracy of the formula is assessed using examples which compute the theoretical signal to noise ratio resolution threshold and compare it with the threshold obtained from simulations.

## 1. Introduction

The problem of estimating the directions of arrival (DOA) of signals produced from narrowband sources and impinging on an array of sensors, may be solved by either using one of the classical methods such as the classical beamformer (with all their well known limitations that have to do, mainly, with the physical size of the array) or by using one of the more succesful high-resolution methods such as MUSIC, Root-MUSIC, Minimum-Norm, ESPRIT, etc. These algorithms may be the final step to a combined procedure or algorithm that uses a preprocessing step, called windowing or prefiltering, and combines it with a classical or high resolution algorithm [1] as the final step of the DOA estimation procedure. The windowing or prefiltering involves the use of rectangular matrices produced in various ways as suggested by numerous investigators (see for example [1], [10]). This leads to a reduced dimension space parameter estimation problem compared to the square matrices of higher dimensions one has to deal with in the case of processing the array data without windowing or prefiltering. Besides the reduced computational burden which is the immediate and obvious benefit, the prefiltering also offers to the estimation procedure more advantages, as explained below.

The use of rectangular prefiltering matrices as preprocessors is an example of a reduced dimension data method applied to the parameter estimation problems. Although covariance matrices of smaller dimensions are used in the final estimation stage, there is an improvement in the signal to noise ratio and an enhancement of the estimating capabilities of some of the applied algorithms [1]. In principle, if the reduced dimension data method is a sufficient statistic no information is lost during the transition from the raw data to the reduced dimension prefiltered data ([2], [3]).

The work presented here extends, in the reduced dimension space, the work of Kaveh and Barabell [4]. In this paper the statistical performance of the MUSIC algorithm preceded by preprocessing with banks of Discrete Prolate Spheroidal Sequences (DPSS's) serving as orthogonal beamformers is investigated. Let N be the number of sensors of an array. Let n be the number of the first n (out of total N) DPSS's. Then the reduced dimension data is the result of operating on the raw data with Nxn ($n \leq N$) weighting matrices $\boldsymbol{W}$ whose columns consist of the n DPSS's. The DPSS's are introduced and studied later.

The bank of n DPSS's is used in order to form a chosen spatial area or angular sector within which lie the angles of arrival of the signals of interest. The operation of the matrix $\boldsymbol{W}$ on the received raw data is a windowing function or spatial filtering which results in filtering out, or at least attenuating as much as possible, interference from unwanted signals (for example, other direct but unwanted signals, multipath etc., which is often the case in applications such as sonar and radar). Also, it reduces the noise and it enhances the signal to noise ratio and, finally, it reduces the dimension of the covariance matrices which in turn means smaller computational burden (for example when $N \geq 16$ then we can have $n \leq 5$). The creation of the spatial sector and the spatial filtering on the raw data that follows, is a preprocessing stage before the stage of the final process (MUSIC in our case) that offers all the previously mentioned advantages. The goal of this work is to verify, through simulations, the validity of our theoretical formula for the resolution threshold. A further step in the future would be the comparison of the two cases (with and without prefiltering).

The case of two narrowband signals with equal powers is considered. Based on the assumption that, after the preprocessing, the initial available information about the signal parameters in an angular sector is retained, the work in [4] is invoked and applied directly in the sector. Since the MUSIC [5] algorithm is used as the final stage of the estimation of the angles of arrival, a new theoretical resolution threshold for the signal to noise ratio is derived analogous to the one given in [4]. This threshold depends on the new smaller dimension n, the angular distance between the directions of arrival of the two signals, the array gain and the number of snapshots K. Theoretical curves of the threshold versus the angular distance are plotted. Our simulations give resolution thresholds that are close to the ones theoretically predicted.

This paper is organized as follows. Section 2 provides a model for the statistical characterization of the source signals and the data, a brief overview of the Discrete Prolate Spheroidal Sequences (DPSS's) and definitions of the Array Beamforming Gain and reduced dimension prefiltered data. In section 3, a brief review of the statistical performance of the MUSIC algorithm for two narrowband sources is given and the accompanying basic equations from related work are cited. The section continues with the extension of this work in the reduced dimension space. In section 4, the formula's theoretical predictions are computed and studied through important examples. For the same important scenarios, these results are compared with the ones produced from their simulated analogs. Finally conclusions are drawn in section 5.

## 2. Model - Definitions

### 2.1 Statistical Characterization of the Model

In this subsection we outline the basic assumptions about the sensor array setup, the statistical character and the properties of the source signals, and the way that all these are interconnected. This is basically the model on which the rest of our analysis will be based.

The adopted model considers a linear array of N omni-directional sensors receiving M (M < N) plane waves with frequency $f_0$, impinging from directions $\theta_1,\ldots,\theta_M$ relative to the broadside of the array. These waves are assumed to be zero mean stationary stochastic processes over the observation interval $T_0$, band-limited to a common frequency band with bandwidth B which may be of the same order of magnitude as the center frequency $f_0$ and they are received in the presence of stationary, zero mean, complex noise. The source signal vector $\boldsymbol{s}(t)$ is denoted by $\boldsymbol{s}(t) = [s_1(t),\ldots,s_M(t)]^T$ where "T" is the matrix transpose. Throughout this paper, small letters in bold font denote vectors while capital letters in bold font denote matrices. During an observation interval $T_0$, K samples of the array output, $\boldsymbol{x}(t)$, each of duration $T_s$, are taken. In the next few paragraphs a generalized model for broadband sources will be developed. For simplicity of explanation, without loss of generality, we limit the analysis to the case of monochromatic plane waves with frequency $f_0$. In the general case of broadband sources, the array output in each interval $T_s$ would be normally decomposed into several frequencies using an FFT.

The signal $x_i(t)$, received at the ith sensor, can be expressed as

$$x_i(t) = \sum_{m=1}^{M} a_{im}\, s_m(t + \tau_{im}) + n_i(t), \qquad (1)$$

where $a_{im}$ is the amplitude response of the ith sensor to the mth source, $\tau_{im}$ is the propagation time difference between the ith sensor and the reference sensor and $n_i(t)$ is the additive noise at the ith sensor.

The observation interval $T_o$ is divided into K non-overlapping snapshot intervals $T_s$. For each of these intervals the array output signals $x_i(f_0)$, at frequency $f_0$, where $i = 1,\ldots, N$, will be given by

$$x_i(f_0) = \sum_{m=1}^{M} a_{im}\, e^{j2\pi f_0 \tau_{im}}\, s_m(f_0) + n_i(f_0), \qquad (2)$$

where $s_m(f_0)$ and $n_i(f_0)$ are the frequency components of $s_m(t)$ and $n_i(t)$ respectively and $j = \sqrt{-1}$. We use the following notation for the received signal, source signal and noise in the frequency domain respectively:

$$\boldsymbol{x}(f_0) \equiv [x_1(f_0),\ldots,x_N(f_0)]^T$$
$$\boldsymbol{s}(f_0) \equiv [s_1(f_0),\ldots,s_M(f_0)]^T$$
$$\boldsymbol{n}(f_0) \equiv [n_1(f_0),\ldots,n_N(f_0)]^T.$$

Based on the above definitions, (2) can be written, in vector-matrix notation, as

$$\boldsymbol{x}(f_0) = \boldsymbol{A}(f_0)\, \boldsymbol{s}(f_0) + \boldsymbol{n}(f_0), \qquad (3)$$

where $\boldsymbol{A}(f_0)$ is the N x M direction matrix at frequency $f_0$ and is given in terms of column vectors,

$$\boldsymbol{A}(f_0) = [\boldsymbol{a}(f_0,\boldsymbol{\tau}_1),\ldots,\boldsymbol{a}(f_0,\boldsymbol{\tau}_M)],$$

where the lth N x 1 vector of delays (phase shifts) $\boldsymbol{a}(f_0,\boldsymbol{\tau}_l)$, used to steer the array beam towards the direction $\theta_l$, is the direction vector at frequency $f_j$, which is given by

$$\boldsymbol{a}(f_0,\boldsymbol{\tau}_l) = [a_{1l}\, e^{j2\pi f_0 \tau_{1l}},\ldots,a_{Nl}\, e^{j2\pi f_0 \tau_{Nl}}]^T. \qquad (4)$$

Since $\boldsymbol{A}(f_0)$ contains information on the unknown parameter vector $\boldsymbol{\theta}=[\theta_1,\ldots,\theta_M]^T$ of the directions of arrival, it can be denoted as $\boldsymbol{A}(f_0,\boldsymbol{\theta})$. For a linear array of omni-directional sensors with the same inter-element spacing d, $\boldsymbol{A}(f_0,\boldsymbol{\theta})$ becomes

$$\boldsymbol{A}(f_0,\boldsymbol{\theta})=\begin{pmatrix} 1 & \cdots & 1 \\ \vdots & \ddots & \vdots \\ e^{j2\pi f_0 (d/c)(N-1)\sin\theta_1} & \cdots & e^{j2\pi f_0 (d/c)(N-1)\sin\theta_M} \end{pmatrix}, \qquad (5)$$

where c is the wave propagation speed and $\boldsymbol{\theta}$ is measured from the axis which is perpendicular to the array endfire. In this case the amplitude response $a_{il}$ is assumed to be one while the propagation time difference $\tau_{il}$ is equal to (d/c) (i-1) $\sin\theta_l$. The lth direction vector is now simplified to

$$\boldsymbol{a}(f_0,\theta_l) = [1,\ldots,e^{j2\pi f_0 (d/c)(N-1)\sin\theta_l}]^T. \qquad (6)$$

Based on the above notation, the spatial covariance matrix $\boldsymbol{R}_x(f_0)$ is given by

$$\boldsymbol{R}_x(f_0) = E[\boldsymbol{x}(f_0)\,\boldsymbol{x}(f_0)^H] = \boldsymbol{A}(f_0,\boldsymbol{\theta})\, E[\boldsymbol{s}(f_0)\,\boldsymbol{s}(f_0)^H]\, \boldsymbol{A}(f_0,\boldsymbol{\theta})^H + E[\boldsymbol{n}(f_0)\,\boldsymbol{n}(f_0)^H], \qquad (7)$$

where 'H' denotes the conjugate transpose. If $T_s$ is sufficiently large at each snapshot, then $\boldsymbol{x}_k(f_0)$, k=1,...,K, can be shown to be approximately uncorrelated (see [6]). Also, $E[\boldsymbol{s}(f_0)\boldsymbol{s}(f_0)^H]=(1/T_s)\boldsymbol{R}_s(f_0)$, where $\boldsymbol{R}_s(f_0)$ is the unknown signal spectral density matrix. By $\boldsymbol{R}_s(f_0)$ we mean the elements of the upper triangular part of the matrix. So (7) becomes

$$\boldsymbol{R}_x(f_0) = \frac{1}{T_s}\,[\boldsymbol{A}(f_0,\boldsymbol{\theta})\boldsymbol{R}_s(f_0)\,\boldsymbol{A}(f_0,\boldsymbol{\theta})^H + \sigma_n^2(f_0)\,\boldsymbol{R}_n(f_0)], \qquad (8)$$

where $\boldsymbol{R}_n(f_0)$ is the noise spectral density matrix and $\sigma_n^2(f_0)$ is the unknown noise spectral power level. Without loss of generality we may assume that $T_s$=1. Finally, the sample spatial covariance matrix $\hat{\boldsymbol{R}}_x(f_0)$ which is an estimate of the true spatial covariance matrix $\boldsymbol{R}_x(f_0)$ is given by

$$\hat{\boldsymbol{R}}_x(f_0) = \frac{1}{K}\sum_{k=1}^{K} \boldsymbol{x}_k(f_0)\,\boldsymbol{x}_k(f_0)^H, \qquad (9)$$

where "^" denotes the sample value or the estimate of an entity.

For a given frequency $f_0$, the vectors $\boldsymbol{x}_k(f_0)$, k=1,...,K, are independent, identically distributed, N-variate, complex, Gaussian random vectors with zero mean and covariance matrix $\boldsymbol{R}_x(f_0)$ given by the equation (9) above.

### 2.2 Discrete Prolate Spheroidal Sequences (DPSS's)

In this section, a few basics about the discrete prolate spheroidal sequences are given. Details can be found in Slepian's work ([7]) where the entire theory is developed and explained analytically. Here are the necessary definitions and notations:

For each $0 \le k \le N-1$, the kth spheroidal sequence $\{\upsilon_n^{(k)},\ 0 \le n \le N-1\}$ with order N and parameter B, consists of the elements of the kth real and normalized eigenvector of the NxN symmetric, positive definite matrix $\boldsymbol{C}$ with (m,n)-th element,

$$C_{mn} = \frac{\sin[2\pi B(m-n)]}{\pi(m-n)}, \qquad 0 \le m \le N-1 \qquad (10)$$

corresponding to the associated eigenvalues $\lambda_k$ in decreasing order ($\lambda_k \le \lambda_{k-1}$). It is known that for any index limited sequence $\{y_n,\ 0 \le n \le N-1\}$ its fractional energy in the band (-B, B) is given as, $E(B) = (\boldsymbol{y}^H \boldsymbol{C}\,\boldsymbol{y})/\,\boldsymbol{y}^H \boldsymbol{y}$, where $\boldsymbol{y} = [y_0, y_1,\ldots, y_{N-1}]$. The maximum for E(B) is reached when $\boldsymbol{y}$ equals the 0th prolate spheroidal sequence with order N and parameter B.

It is known from [7] that as $N\to\infty$: (i) $\lambda_k\to 1$ if $k=2BN(1-\varepsilon)$ and (ii) if $k=2BN(1+\varepsilon)$, $\lambda_k\to 0$. This is true for any $\varepsilon=\varepsilon(N)$ satisfying $0<\varepsilon<1$. Thus a fraction arbitrarily close to 2B of the bandlimited DPSS's are confined almost entirely to the index set $0 \le n \le N-1$. The remaining DPSS's have almost none of their energy in this index set. In practice, if $k \le [2BN] = n$, $\lambda_k$ is very close to one. Thus, the first n prolate spheroidal sequences provide a set of orthonormal sequences that have most of their energy concentrated in the spatial frequency band (-B, B).

### 2.3 Reduced Data Definition

In this section we present the definitions of the weighting matrix and the array beamforming gain. We define the Nxn weighting matrix $\boldsymbol{W}(f_0,\theta_0)$ with elements

$$w_{lk} = \varepsilon_k v_l^{(k)}\, e^{j2\pi f_0 \tau_l(\theta_0)}, \qquad (11)$$

where $v^{(k)}$, $0 \le l \le N-1$, $0 \le k \le n-1$, is the l-th element of the k-th prolate spheroidal sequence with order N and parameter B and $\theta_0$ is the spatial center of the sector $[\theta_0-\Theta\ \ \theta_0+\Theta]$. The symbol $\varepsilon_K$ is 1 for k even and -j ($j=\sqrt{-1}$) for k odd and is used as a normalization constant. The nx1 ($n \le N$) reduced dimension data vectors $\boldsymbol{y}_k(f_0)$, k=1,...,K are given by

$$\boldsymbol{y}_k(f_0) = \boldsymbol{W}(f_0,\theta_0)^H \boldsymbol{x}_k(f_0), \qquad k=1,...,K. \qquad (12)$$

The data vectors $\boldsymbol{y}_k(f_0)$ are the result of a linear operation onto the raw data vectors $\boldsymbol{x}_k(f_0)$. Therefore, they also, are independent, identically distributed, n-variate, complex normal random vectors with zero mean and covariance matrix

$$\boldsymbol{R}_y(f_0)= \frac{1}{T_s} [\boldsymbol{W}(f_0,\theta_0)^H \boldsymbol{A}(f_0,\boldsymbol{\theta})\boldsymbol{R}_s(f_0)\boldsymbol{A}(f_0, \boldsymbol{\theta})^H \boldsymbol{W}(f_0,\theta_0)$$
$$+ \sigma_n^2(f_0)\boldsymbol{W}(f_0,\theta_0)^H \boldsymbol{R}_n\boldsymbol{W}(f_0,\theta_0)]. \qquad (13)$$

The Array beamforming Gain $A_g$ which is generally equal to $10\log_{10}||\boldsymbol{a}(f_0,\theta)||^2/N$ (see [8]) may be modified in the reduced dimension space as follows:

$$A_g = 10\log_{10} [||\boldsymbol{W}(f_0,\theta_0)^H \boldsymbol{a}(f_0,\theta)||^2/N], \qquad (14)$$

where $\boldsymbol{a}(f_0,\theta)$ has been defined in (6).

## 3. Statistical Performance of the MUSIC Algorithm in Angular Sectors

The performance of the various well known and successful eigen-decomposition methods has been theoretically studied. The statistical performance of MUSIC, Minimum-Norm and Maximum Likelihood algorithms, has been studied in [4] [9], [10], [11] and [12]. The statistical performance studies have been extended with the analysis of ROOT-MUSIC method for linear uniform arrays in [13] and [14].

In the following sections we will study theoretically the performance of prefiltered MUSIC. Firstly, a review of past results will be given and based on these results, the theoretical resolution threshold in the case of Prefiltered MUSIC will be obtained and will be verified through simulations.

### 3.1. Review of the Resolution Threshold of the MUSIC Algorithm

Firstly, we review notations and results from [4] which will help us formulate our method. We assume that all of the quantities will be functions of the radian frequency $\boldsymbol{\omega}$ instead of cyclical frequency $f_0$. Also, since the analysis will be concerned with two narrowband sources which are assumed to be functions of a single frequency, for simplicity, the data vectors will not be written as functions of frequency. So, for M monochromatic plane waves of frequency $f_0$ impinging from directions $\theta_1,...,\theta_M$, relative to the broadside of the array, forming the direction vector $\boldsymbol{\theta}=[\theta_1,...,\theta_M]^T$ of the directions of arrival, we define the vector $\boldsymbol{\omega}=[\omega_1,..., \omega_M]^T$ where $\omega_i=2\pi f_0(d/c)\sin\theta_i$ and d, c, K have been defined in section II. The direction matrix $\boldsymbol{A}$ is then a function of $\boldsymbol{\omega}$ i.e. $\boldsymbol{A}(\boldsymbol{\omega})$, so according to the above setup, the sample covariance matrix $\boldsymbol{R}_x$ in (8) is written as

$$\boldsymbol{R}_x = E[\boldsymbol{x}_k \boldsymbol{x}_k^H] = \boldsymbol{A}(\boldsymbol{\omega}) \boldsymbol{R}_s \boldsymbol{A}(\boldsymbol{\omega})^H + \sigma_n^2\boldsymbol{I}, \qquad (15)$$

where $\boldsymbol{R}_s$ is the signal power spectral density matrix while the estimate $\hat{\boldsymbol{R}}_x$ (based on the available data) of $\boldsymbol{R}_x$ is written as

$$\hat{\boldsymbol{R}}_x = \frac{1}{K}\sum_{k=1}^{K} \boldsymbol{x}_k \boldsymbol{x}_k^H. \qquad (16)$$

The above matrix is the statistic on which the angular spectral estimates of $\omega_i$ are based. The MUSIC algorithm is based on the decomposition of the matrix $\boldsymbol{R}_x$ as

$$\boldsymbol{R}_x = \sum_{i=1}^{N} \lambda_i \boldsymbol{e}_i \boldsymbol{e}_i^H , \qquad (17)$$

where $\lambda_1 \ge \lambda_2 \ge ... \ge \lambda_M > \lambda_{M+1} = ... = \lambda_N = \sigma_n^2$ are the eigenvalues of $\boldsymbol{R}_x$ and $e_i$, i=1,...,N are its orthonormal eigenvectors. According to the MUSIC algorithm the values $\omega_i$, of $\omega$ (where $\omega$ is an independent scalar variable) for which the so called spectrum $P(\omega_i)$ becomes infinity, or if

$$P(\omega) = \frac{1}{D(\mu)}, \qquad (18)$$

the values $\omega_i$ of $\omega$ for which $D(\omega_i) = 0$, are the exact estimates of the parameters $\omega_i$. The spectrum $D(\omega)$ is called null spectrum and is given alternatively in terms of noise -and - signal subspace by (see[15])

$$D(\omega) = \boldsymbol{a}(\omega)^H [\sum_{i=M+1}^{N} \boldsymbol{e}_i \boldsymbol{e}_i^H ] \boldsymbol{a}(\omega) \qquad (19)$$

or

$$D(\omega) = \boldsymbol{a}(\omega)^H [\boldsymbol{I} - \sum_{i=1}^{M} \boldsymbol{e}_i \boldsymbol{e}_i^H ] \boldsymbol{a}(\omega), \qquad (20)$$

where $\boldsymbol{I}$ is the unitary matrix. Since $\boldsymbol{a}(\omega)^H \boldsymbol{a}(\omega) =1$, $D(\omega)$ becomes,

$$D(\omega) =1 - \boldsymbol{a}(\omega)^H (\sum_{i=1}^{M} \boldsymbol{e}_i \boldsymbol{e}_i^H ) \boldsymbol{a}(\omega). \qquad (21)$$

Since the exact covariance matrix $\boldsymbol{R}_x$ is not known and only its estimate, $\hat{\boldsymbol{R}}_x$ is available through sampling, the performance of the MUSIC algorithm may be analyzed only in terms of the approximate statistical behavior of the estimated null spectrum $\hat{D}(\omega)$ which is given by,

$$\hat{D}(\omega) = 1 - \boldsymbol{a}(\omega)^H [\sum_{i=1}^{M} \hat{\boldsymbol{e}}_i \hat{\boldsymbol{e}}_i^H] \boldsymbol{a}(\omega).$$

Let the estimated values of the eigenvectors and eigenvalues be $\hat{\boldsymbol{e}}_i = \boldsymbol{e}_i + \boldsymbol{\eta}_i$, $\hat{a}_i = \boldsymbol{\lambda}_i + \beta_i$ where $\boldsymbol{\eta}_i$ and $\beta_i$ are the errors of the estimated eigenvectors and eigenvalues respectively. Following a similar analysis as in Kaveh and Barabell in [4] the expected value of $\hat{D}(\omega)$ is given by,

$$E[\hat{D}(\omega)] \approx 1-\boldsymbol{a}(\omega)^H[\sum_{i=1}^{M} \boldsymbol{e}_i\boldsymbol{e}_i^H]\boldsymbol{a}(\omega)-\boldsymbol{a}(\omega)^H[\sum_{i=1}^{M} \boldsymbol{\eta}_i\boldsymbol{\eta}_i^H]\boldsymbol{a}(\omega)$$

$$- 2 \,\mathrm{Re}[\boldsymbol{a}(\omega)^H [\sum_{i=1}^{M} \boldsymbol{e}_i \overline{\boldsymbol{\eta}}_i^H ] \boldsymbol{a}(\omega)].$$

The bias and variance of $\hat{D}(\omega)$, especially in the neighborhood of $\omega_i$, can be interpreted as indicators of the resolving capabilities of MUSIC. Using the estimation approach in [4] for the expected value of $\boldsymbol{\eta}_i$, the estimated null spectrum is given by

$$E[\hat{D}(\omega)] \approx D(\omega) - \boldsymbol{a}(\omega)^H [\sum_{i=1}^{M} \sum_{\substack{j=1\\ j\neq i}}^{N} \frac{a_i a_j}{K(a_i - a_j)^2}$$

$$(\boldsymbol{e}_j \boldsymbol{e}_j^H - \boldsymbol{e}_i \boldsymbol{e}_i^H)] \boldsymbol{a}(\omega). \qquad (22)$$

In the case of two signals the above expected value at frequency $\omega_k$ is given by

$$E[\hat{D}(\omega_k)] \approx \frac{(N-2)}{K} [\frac{a_1 \pm_n^2}{(a_1 - \pm_n^2)^2} | \boldsymbol{a}(\omega_k)^H \boldsymbol{e}_1|^2$$

$$+ \frac{a_2 \pm_n^2}{(a_2 - \pm_n^2)^2} | \boldsymbol{a}(\omega_k)^H \boldsymbol{e}_2|^2 ] \qquad (23)$$

and its variance

$$\mathrm{Var}[\hat{D}(\omega_k)] \approx \frac{2\pm_n^2 (N-2)}{K} \{\sum_{i=1}^{2} \frac{a_i \pm_n^2}{(a_i - \pm_n^2)^2}$$

$$[| \boldsymbol{a}(\omega_k)^H \boldsymbol{e}_i|^2 - | \boldsymbol{a}(\omega_k)^H \boldsymbol{e}_i|^4]$$

$$- \frac{a_i \pm_n^2}{(a_i - \pm_n^2)^2} | \boldsymbol{a}(\omega_k)^H \boldsymbol{e}_i \boldsymbol{e}_i^H \boldsymbol{a}(\omega_k)|^2\}. \qquad (24)$$

For equal power signals, $E[\hat{D}(\omega_k)]$ and $E[\hat{D}(\omega_m)]$ are approximated, in terms of the array signal to noise ratio $\xi = \frac{NP}{\pm_n^2}$, the number of sensors N, the number of samples K, and the angular separation $\Delta = \frac{N(\mu_1 - \mu_2)}{2\sqrt{3}}$, as

$$E[\hat{D}(\omega_k)] \approx \frac{(N-2)}{K} [\frac{1}{¶} + \frac{1}{¶^2 \Delta^2}], \qquad (25)$$

$$E[\hat{D}(\omega_m)] \approx \frac{\Delta^4}{80} + \frac{(N-2)}{K} [\frac{4+\Delta^2}{8¶} + \frac{2\Delta^2 + \Delta^4}{8¶^2\Delta^2}], \quad (26)$$

where $\omega_m = \frac{\mu_1 + \mu_2}{2}$. The two signals are considered to be resolved at the signal to noise ratio at which $E[\hat{D}(\omega_1)] \le E[\hat{D}(\omega_2)] \le E[\hat{D}(\omega_m)]$. The reasons for this conjecture are the following. Resolution is achieved when $\hat{D}(\omega_1)$ and $\hat{D}(\omega_2)$ are both less than $\hat{D}(\omega_m)$ which means that $\hat{P}(\omega_1)$ and $\hat{P}(\omega_2)$ are both greater than $\hat{P}(\omega_m)$. When the above equality is true the probability of resolution ranges from, approximately, 0.33 when the variations of $\hat{D}(\omega_1)$, $\hat{D}(\omega_2)$ and $\hat{D}(\omega_m)$ are totally independent, to nearly 0.5 for the situation when $\hat{D}(\omega_1)$ and $\hat{D}(\omega_2)$ are completely correlated. After equating the right sides of (25) and (26) the threshold $\xi_T$ for which $E[\hat{D}(\omega_1)] \le E[\hat{D}(\omega_2)] \le E[\hat{D}(\omega_m)]$ is given by

$$\xi_T = \frac{1}{K} \{20(N-2)\Delta^{-4}[1+ \sqrt{1+\frac{K}{5(N-2)}\Delta^2} ]\}. \qquad (27)$$

Some observations about $\xi_T$ may be made immediately. For large N, $\xi_T$ is approximately proportional to N. Also, for $K \ll 5N\Delta^{-2}$, $\xi_T$ varies as $K^{-1}\Delta^{-4}$ while for $5N\Delta^{-2} \ll K$, it varies as $\Delta^{-3}N^{-1/2}$.

### 3.2. Resolution Threshold for the Prefiltered MUSIC

The nx1 prefiltered data vector $\boldsymbol{y}_k$, used in various places in the previous section, is given by

$$\boldsymbol{y}_k = \boldsymbol{W}(f_0,\theta_0)^H \boldsymbol{x}_k, \qquad k=1,...,K, \tag{28}$$

where the elements of $\boldsymbol{W}(f_0,\theta_0)^H$ have been defined in (11) and $\theta_0$ is the center of the angular sector outside which we want all signals to be attenuated. The vectors $\boldsymbol{y}_k$, k=1,...,K, are complex, zero mean, circular Gaussian vectors. The respective nxn covariance matrix is given by

$$\boldsymbol{R}_y = \boldsymbol{W}(f_0,\theta_0)^H \boldsymbol{A}(\omega)\, \boldsymbol{R}_s\, \boldsymbol{A}(\omega)^H \boldsymbol{W}(f_0,\theta_0) + \sigma_n^2(f_0)\boldsymbol{I}_n, \tag{29}$$

where $\boldsymbol{A}(\omega)$ is defined in equations (5) and (6). The matrix $\boldsymbol{R}_y$ is decomposed as

$$\boldsymbol{R}_y = \sum_{i=1}^{n} \tilde{a}_i\, \boldsymbol{e}_i^y\, \boldsymbol{e}_i^{yH}, \tag{30}$$

where $\tilde{a}_1 \geq \tilde{a}_2 \geq ... \geq \tilde{a}_M > \tilde{a}_{M+1} = ... = \tilde{a}_N = \sigma_n^2$ are the eigenvalues of $\boldsymbol{R}_y$ and $\boldsymbol{e}_i^y$, i=1,...,n are its orthonormal eigenvectors. The prefiltered spectrum is written as

$$P_y(\omega) = \frac{1}{D_y(\mu)}, \tag{31}$$

where $D_y(\omega)$ is given by

$$D_y(\omega)=\boldsymbol{a}(\omega)^H \boldsymbol{W}(f_0,\theta_0)[\sum_{i=M+1}^{n} \boldsymbol{e}_i^y\, \boldsymbol{e}_i^{yH}]\boldsymbol{W}(f_0,\theta_0)^H \boldsymbol{a}(\omega). \tag{32}$$

Following the reasoning in equations (20-21) above, this can be expressed also as

$$D_y(\omega)=1-\boldsymbol{a}(\omega)^H \boldsymbol{W}(f_0,\theta_0)[\sum_{i=1}^{M} \boldsymbol{e}_i^y\, \boldsymbol{e}_i^{yH}]\boldsymbol{W}(f_0,\theta_0)^H \boldsymbol{a}(\omega). \tag{33}$$

The estimated null spectrum $\hat{D}_y(\omega)$ is written as

$$\hat{D}_y(\omega)=1-\boldsymbol{a}(\omega)^H \boldsymbol{W}(f_0,\theta_0)[\sum_{i=1}^{M} \hat{\boldsymbol{e}}_i^y\, \hat{\boldsymbol{e}}_i^{yH}]\boldsymbol{W}(f_0,\theta_0)^H \boldsymbol{a}(\omega). \tag{34}$$

Its expected value (analogously to equation (22)) is approximated as follows:

$$E[\hat{D}_y(\omega)] \approx D_y(\omega)$$

$$- \boldsymbol{a}(\omega)^H \boldsymbol{W}(f_0,\theta_0)[\sum_{i=1}^{M} \sum_{j=1}^{N} \frac{\tilde{a}_i \tilde{a}_j}{K(\tilde{a}_i - \tilde{a}_j)^2}]\boldsymbol{W}(f_0,\theta_0)^H \boldsymbol{a}(\omega). \tag{35}$$

For two signals, the expected value and variance at spatial frequency $\omega_k$ may be approximated, in the sense of [4] as

$$E[\hat{D}_y(\omega_k)] \approx \frac{(n-2)}{K} [\frac{\tilde{a}_1 \pm_n^2}{(\tilde{a}_1 - \pm_n^2)^2} |\boldsymbol{a}(\omega_k)^H \boldsymbol{W}(f_0,\theta_0)\, \boldsymbol{e}_1^y|^2$$

$$+ \frac{\tilde{a}_2 \pm_n^2}{(\tilde{a}_2 - \pm_n^2)^2} |\boldsymbol{a}(\omega_k)^H \boldsymbol{W}(f_0,\theta_0)\, \boldsymbol{e}_2^y|^2], \tag{36}$$

$$\mathrm{Var}[\hat{D}_y(\omega_k)] \approx \frac{\pm_n^2 (n-2)}{K} \{\sum_{i=1}^{2} \frac{\tilde{a}_i \pm_n^2}{(\tilde{a}_i - \pm_n^2)^2}$$

$$[|\boldsymbol{a}(\omega_k)^H \boldsymbol{W}(f_0,\theta_0)\boldsymbol{e}_i^y|^2 - |\boldsymbol{a}(\omega_k)^H \boldsymbol{W}(f_0,\theta_0)\, \boldsymbol{e}_i^y|^4]$$

$$-\sum_{i=1}^{2} \frac{\tilde{a}_i \pm_n^2}{(\tilde{a}_i - \pm_n^2)^2} |\boldsymbol{a}(\omega_k)^H \boldsymbol{W}(f_0,\theta_0)\boldsymbol{e}_i^y \boldsymbol{e}_i^{yH} \boldsymbol{W}(f_0,\theta_0)^H \boldsymbol{a}(\omega_k)|^2\}. \tag{37}$$

In order to obtain simplified expressions for $E[\hat{D}_y(\omega_k)]$ and $\mathrm{Var}[\hat{D}_y(\omega_k)]$ in terms of n, K, Δ and the array signal to noise ratio¶ after the prefiltering, $|\boldsymbol{a}(\omega_k)^H \boldsymbol{W}(f_0,\theta_0)\boldsymbol{e}_i^y|$ and $|\boldsymbol{a}(\omega_m)^H \boldsymbol{W}(f_0,\theta_0)\boldsymbol{e}_i^y|$ need to be approximated. We next give a method for such an approximation.

We assume that the two uncorrelated signals in the angular sector have powers $\tilde{P}_i$, i=1,2, after the prefiltering. These powers are connected to the initial signal power $P_i$ with the relation $\tilde{P}_i = A_g(\omega_i)\, P_i$, where $A_g(\omega_i)$ is the array gain for the signal at $\omega$, and is defined in (14). Based on this notation, the eigenvalues are given by (see [16])

$$\tilde{a}_{1(2)} = \frac{1}{2}(\tilde{P}_1 + \tilde{P}_2)N[1+\sqrt{1+\frac{4\tilde{P}_1\tilde{P}_2(1-|\tilde{\Phi}|^2)}{(\tilde{P}_1+\tilde{P}_2)}}]+\sigma_n^2. \tag{38}$$

Let the eigenvectors be unnormalized, $\tilde{a}_1' = \tilde{a}_1 - \sigma_n^2$, and let $\tilde{\Phi}$ denote the cosine of the angle between the vectors $\boldsymbol{a}(\omega_1)^H \boldsymbol{W}(f_0,\theta_0)$ and $\boldsymbol{a}(\omega_2)^H \boldsymbol{W}(f_0,\theta_0)$ given by the fraction

$$\tilde{\Phi} = \frac{[a(\mu_1)^H W(f_0,\bar{}_0) W(f_0,\bar{}_0)^H a(\mu_2)]}{[\|a(\mu_1)^H W(f_0,\bar{}_0)\| \|a(\mu_2)^H W(f_0,\bar{}_0)\|]}. \tag{39}$$

Then we have

$$e_l^y \propto a(\omega_1)^H W(f_0,\theta_0) + \frac{\tilde{a}_1' - \tilde{P}_1}{\tilde{P}_1 |\tilde{\Phi}|} a(\omega_2)^H W(f_0,\theta_0), \quad (40)$$

where $\propto$ is used to indicate that the expression is not normalized. It is difficult to approximate $\tilde{\Phi}$ directly but, we may assume that it is very close to the cosine $\Phi$ of the angle between the direction vectors $a(\omega_1)$ and $a(\omega_2)$. For two closely spaced signals, $\Phi$ can be written, in terms of $\omega_d = \frac{\mu_1 - \mu_2}{2}$ as,

$$\Phi = a(\omega_1)^H a(\omega_2) = \frac{1}{N} e^{-j(N-1)\mu_d} \frac{\sin(N\mu_d)}{\sin(\mu_d)}. \quad (41)$$

Let us assume that $(N\,\omega_d)^2 \ll 1$ then $\Phi$ and consequently $\tilde{\Phi}$, may be expanded as

$$\tilde{\Phi} \approx \Phi e^{-j(N-1)\mu_d} [1 - \frac{1}{6} N^2 \omega_d^2 + \frac{1}{120} N^4 \omega_d^4 - \dots]. \quad (42)$$

Now if $\Delta^2 = \frac{N^2 \mu_d^2}{3}$ then the above approximation becomes

$$|\tilde{\Phi}| \approx |\Phi| \approx 1 - \frac{\Delta^2}{2} + \frac{9}{120}\Delta^4 - \dots \quad (43)$$

Let us assume that $\tilde{P}_1 = \tilde{P}_2 = \tilde{P}$, and an angular sector center $\theta_0$ placed between the directions of arrival of the two signals, then $A_g(\omega_1) \approx A_g(\omega_2) \approx A_g$ and equations B(7), B(8), B(9) and B(10) in [4] may be modified, in the case of prefiltering as follows:

$$|a(\omega_i)^H W(f_0,\theta_0)\, e_1^y|^2 = \frac{\tilde{a}_1'}{2N\tilde{P}} = \frac{\tilde{a}_1'}{2NA_g\tilde{P}} \approx 1 - \frac{\Delta^2}{4} + \frac{3}{80}\Delta^4,$$

for i=1,2 (44)

$$|a(\omega_i)^H W(f_0,\theta_0)\, e_2^y|^2 = \frac{\tilde{a}_2'}{2N\tilde{P}} = \frac{\tilde{a}_2'}{2NA_g\tilde{P}} \approx \frac{\Delta^2}{4} - \frac{3}{80}\Delta^4,$$

for i=1,2 (45)

and

$$|a(\omega_m)^H W(f_0,\theta_0)\, e_1^y|^2 \approx 1 - \frac{1}{80}\Delta^4, \quad (46)$$

$$|a(\omega_m)^H W(f_0,\theta_0)\, e_2^y|^2 \approx 0. \quad (47)$$

From Eqns (34) and (35) and based on the approximations in Eqns (44) through (47), $E[\hat{D}_y(\omega_\kappa)]$ and $E[\hat{D}_y(\omega_m)]$ will be given by

$$E[\hat{D}_y(\omega_\kappa)] \approx \frac{(N-2)}{K} [\frac{1}{2\tilde{¶}} + \frac{1}{\tilde{¶}^2 \Delta^2}], \quad (48)$$

$$E[\hat{D}_y(\omega_m)] \approx \frac{\Delta^4}{80} + \frac{(N-2)}{K} [\frac{4+\Delta^2}{8\tilde{¶}} + \frac{2\Delta^2 + \Delta^4}{8\tilde{¶}^2 \Delta^2}], \quad (49)$$

where $\tilde{¶} = A_g \frac{NP}{\pm_n^2} = A_g\, \xi_T$. The term $\xi_T$ is the theoretical array signal to noise ratio (ASNR) after the prefiltering and the dimension reduction. By equating the right parts of the two above equations and solving for $\tilde{¶}$ we find that the resolution threshold $\xi_T$ in the original space, in terms of the reduced dimension n, is given by

$$\xi_T = \frac{1}{A_g K} \{20(n-2)\Delta^{-4} [1 + \sqrt{1 + \frac{K}{5(n-2)}\Delta^2}\,]\}. \quad (50)$$

The above expression is analogous to the one obtained in [4]. Comparing (27) and (50) we see that they are similar except that the initial number of sensors N, in (27), has been replaced by the reduced dimension n and, also, the array gain $A_g$ has been added in (50). The array gain which was defined in (14) is used in (50) and in the results as a dimensionless number. It is noted that the angular separation $\Delta$ in the reduced dimension space, remains the same as in the initial element space.

## 4. Simulation Results

In order to verify our results, we selected the case of two linear arrays with different numbers of equally spaced sensors. We also tested our method for several scenarios of different beamwidths with increasing values. In order to verify the accuracy of the theoretical results, two sets of simulations were carried out using the model described in section II. In the first set a linear uniform array of N=8 sensors is used and in the second set a linear uniform array of N=16 sensors is used. In both cases the arrays receive two equi-powered signals impinging from directions $\theta_1$ and $\theta_2 = -\theta_1$ and having wavelengths equal to twice the array inter-element distance. The term $\Delta$ which is used to compute the theoretical resolution threshold is expressed as,

$$\Delta = \frac{N(\mu_1 - \mu_2)}{2\sqrt{3}} = \frac{N\text{®}[\sin(\theta_1) - \sin(\theta_2)]}{2\sqrt{3}} \approx \frac{N\text{®}\sin(\theta_d/2)}{\sqrt{3}},$$

where the angle separation $\theta_d=2\theta_1$. The columns in the weighting matrix $\boldsymbol{W}$ consist of n=3 DPSS's which are computed for B=0.0781, when N=8 and N=16 while the sector center $\theta_0$ is chosen to be $\theta_0=(\theta_1+\theta_2)/2$. In all cases, the two signals were considered to be resolved if two spectral peaks were found within a spatial area equal to a beamwidth around the midpoint between the true directions of arrival. The simulated resolution threshold $\xi_S$ for the various examples was determined as follows. For a set of pairs of equal signal power values, which were selected in an increasing order with a step of 1dB, 30 simulated runs were performed for each set. The number of times (runs) during which both signals were resolved (according to the definition above) was counted and compared against the total number of runs (30) and thus the probability of resolution was computed. The minimum value of the signal to noise ratio for which the probability of resolution was unity, is the threshold $\xi_S$.

Tables 1 and 2 show comparisons for various angle separations and numbers of snapshots, between the theoretical element resolution threshold ($\xi_T$) and the threshold ($\xi_S$) obtained through simulations . We can observe that in all cases as the number of snapshots increases (increased received information) the threshold drops while the difference between theoretical and simulated results decreases and they become, approximately, the same. Also as $\theta_d$ increases, the resolution threshold drops to very small values, as expected. Figures 1 and 2 show the ASNR term $\xi_T$ calculated, in decibels (dB), for various important scenarios of sensors' numbers N, numbers n of DPSS's and numbers of snapshots K and is plotted versus the angle separation ($\theta_d$) of the two signals. The results in the figures illustrate the theoretical part of the results in the tables. Therefore we conclude that the method performs very accurately.

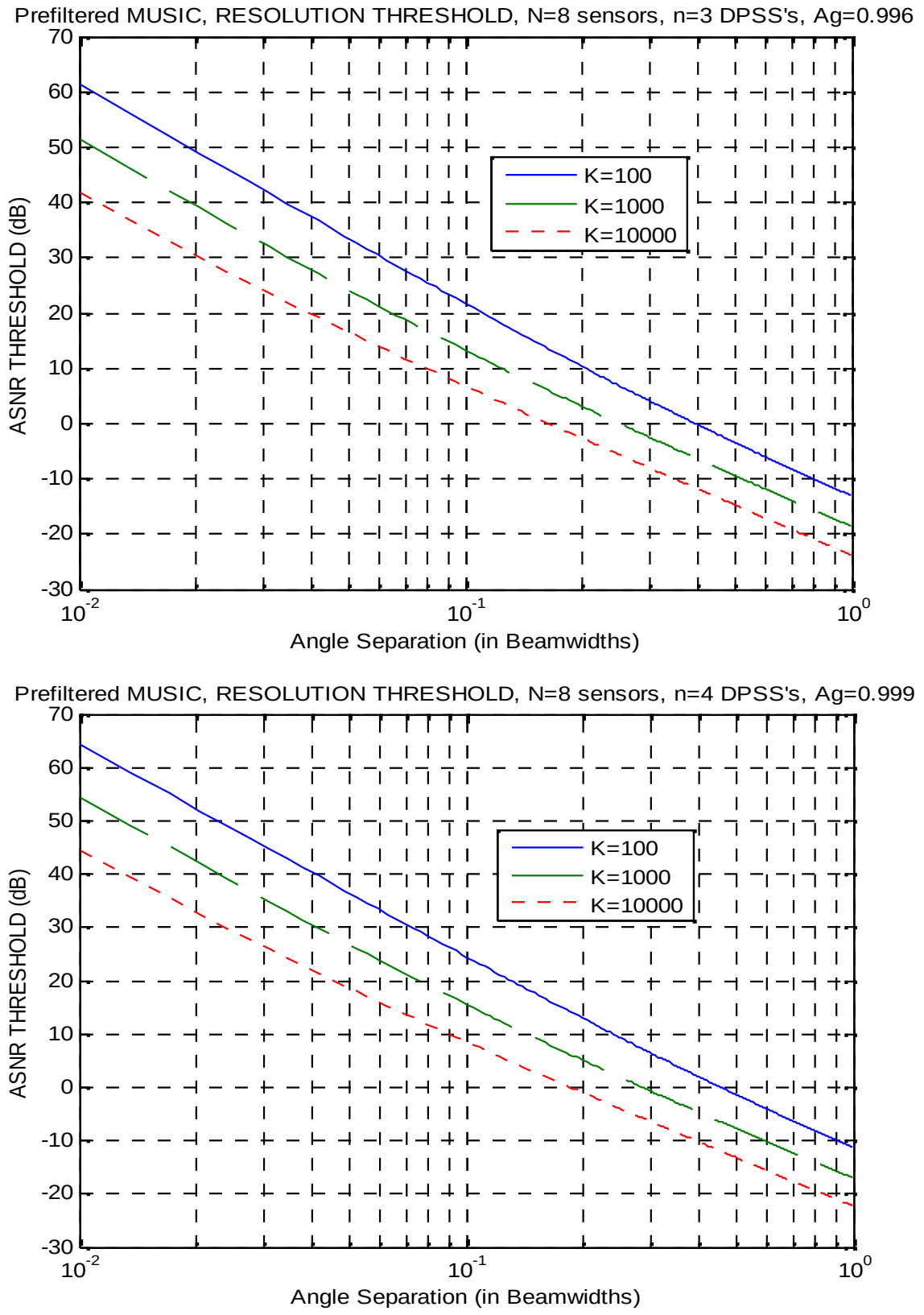

Fig. 1. Theoretical Array Signal to Noise Ratio performance of Prefiltered MUSIC for N=8 sensors and various numbers of DPSS's

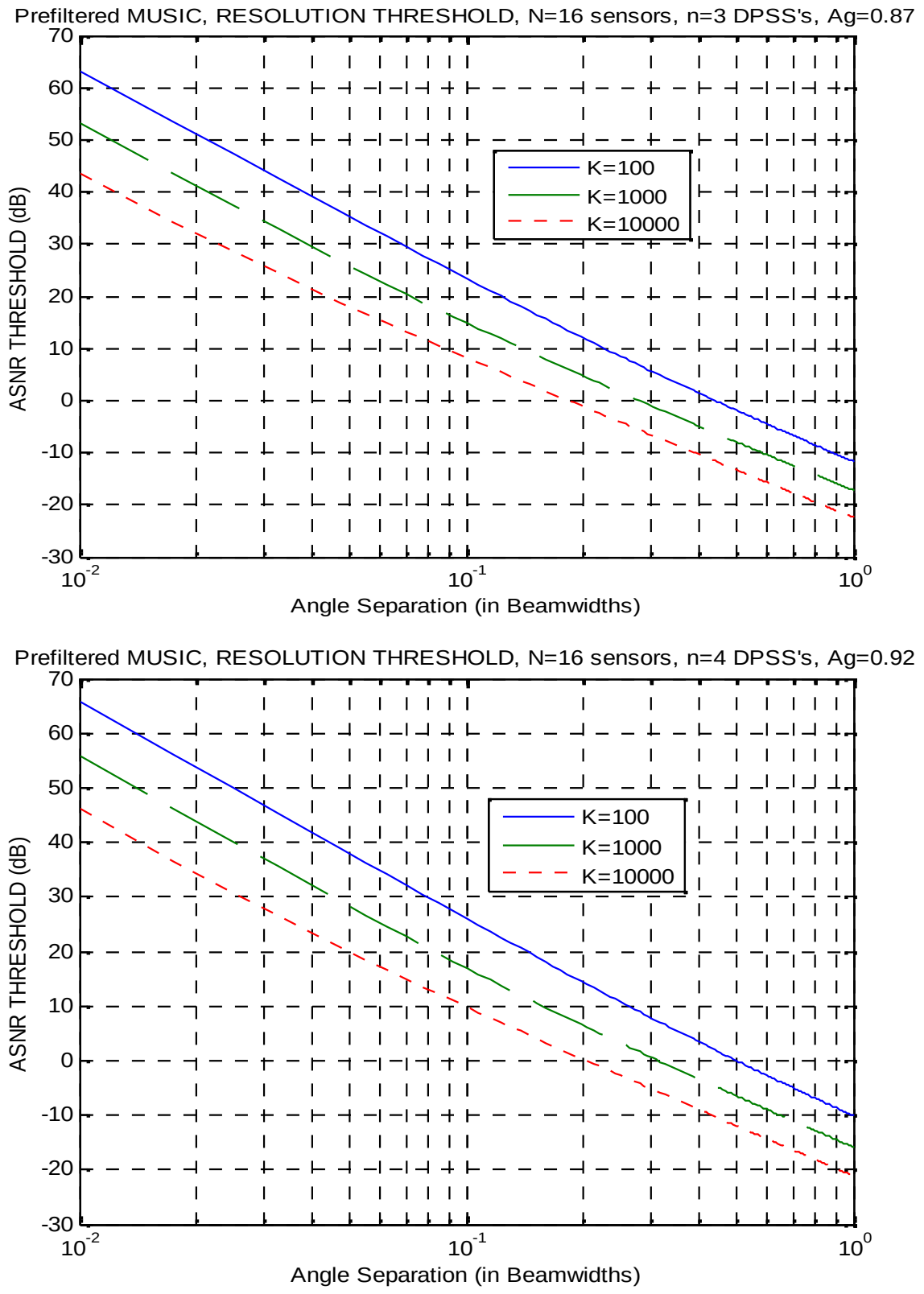

Fig. 2. Theoretical Array Signal to Noise Ratio performance of Prefiltered MUSIC for N=16 sensors and various numbers of DPSS's

TABLE 1
Theoretical Array Signal to Noise Ratio Threshold for Prefiltered MUSIC

| $N$=8 sensors, $\theta_d$ = 0.8° = 0.05 Beamwidths | | |
|---|---|---|
| # of snapshots | $\xi_T$(Theory) (dB) | $\xi_S$(Simul.) (dB) |
| 100 | 36 | 37 |
| 1000 | 27.19 | 27 |
| 10,000 | 20.32 | 19 |
| $N$=8 sensors, $\theta_d$ = 2° = 0.122 Beamwidths | | |
| # of snapshots | $\xi_T$(Theory) (dB) | $\xi_S$(Simul.) (dB) |
| 100 | 20.88 | 23 |
| 1000 | 13.62 | 15 |
| 10,000 | 7.81 | 8 |
| $N$=8 sensors, $\theta_d$ = 4° = 0.25 Beamwidths | | |
| # of snapshots | $\xi_T$(Theory) (dB) | $\xi_S$(Simul.) (dB) |
| 100 | 10.26 | 12 |
| 1000 | 4 | 6 |
| 10,000 | -1.41 | -1 |
| $N$=8 sensors, $\theta_d$ = 13° = 0.8 Beamwidths | | |
| # of snapshots | $\xi_T$(Theory) (dB) | $\xi_S$(Simul.) (dB) |
| 100 | -6.34 | -3 |
| 1000 | -11.74 | -8 |
| 10,000 | -16.87 | -13 |

TABLE 2
Theoretical Array Signal to Noise RatioThreshold for Prefiltered MUSIC

| $N$=16 sensors, $\theta_d$ = 0.4° = 0.05 Beamwidths | | |
|---|---|---|
| # of snapshots | $\xi_T$(Theory) (dB) | $\xi_S$(Simul.) (dB) |
| 100 | 36.61 | 38 |
| 1000 | 27.78 | 30 |
| 10,000 | 20.90 | 21 |
| $N$=16 sensors, $\theta_d$ = 1° = 0.13 Beamwidths | | |
| # of snapshots | $\xi_T$(Theory) (dB) | $\xi_S$(Simul.) (dB) |
| 100 | 21.47 | 23 |
| 1000 | 14.21 | 16 |
| 10,000 | 8.40 | 8 |
| $N$=16 sensors, $\theta_d$ = 2° = 0.25 Beamwidths | | |
| # of snapshots | $\xi_T$(Theory) (dB) | $\xi_S$(Simul.) (dB) |
| 100 | 10.84 | 14 |
| 1000 | 4.59 | 6 |
| 10,000 | -0.83 | 1 |
| $N$=16 sensors, $\theta_d$ = 6° = 0.8 Beamwidths | | |
| # of snapshots | $\xi_T$(Theory) (dB) | $\xi_S$ (Simul.) (dB) |
| 100 | -4.69 | -1 |
| 1000 | -10.12 | -7 |
| 10,000 | -15.26 | -12 |

## 5. Conclusions

In this paper, we reviewed the past results for the resolution threshold of the MUSIC algorithm for two equi-powered, uncorrelated, narrowband signals. We then derived a formula which gives a theoretical array signal to noise ratio resolution threshold when the MUSIC algorithm is applied after a dimension reducing preprocessing step. The newly derived formula also predicts higher resolution threshold as the number of columns of the prefiltering matrix $\boldsymbol{W}$ increases. We demonstrated by simulations that the method gives accurate results.

## Acknowledgements

This research of A. T. Chronopoulos was partly supported by a NSF grant (HRD-0932339) to the University of Texas at San Antonio. We appreciate the reviewers' comments that enhanced the quality of presentation of our paper.

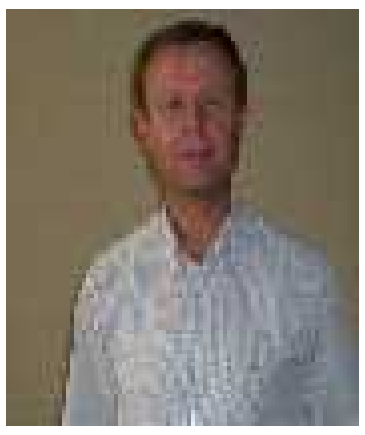

Antonios Bassias holds a Diploma, a MS and a Ph.D degrees, all in Electrical Engineering. He is currently working as a Telecommunications Specialist for the Hellenic Telecommunications Organization (OTE) in Greece. His research interests include digital and wireless communications, array signal processing, underwater communications and underwater signal processing.

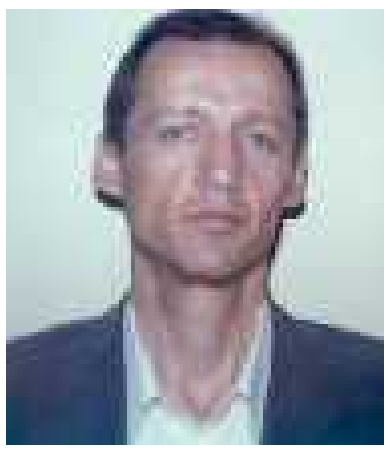

Anthony T. Chronopoulos received his Ph.D. in Computer Science from the University of Illinois at Urbana-Champaign in 1987. He is currently a Professor in Computer Science at the University of Texas at San Antonio. He has published 45 journal and 59 peer-reviewed conference proceedings publications in the areas of Communications, Distributed and Parallel Computing, High Performance Computing, Scientific Computing. He has been awarded 15 federal/state government research grants. His work is cited in over 400 non-coauthors' research articles. He is a senior member of IEEE (since 1997).